\documentclass[aps,prb,preprint,showpacs,groupedaddress]{revtex4}  

\usepackage{graphicx}
\usepackage{dcolumn}
\usepackage{bm}
\usepackage{amsmath}
\usepackage{amssymb}

\newcommand{\vect}[1]{\mbox{{\boldmath $#1$}}}

\newcommand{\vd}[2]{\mbox{$\vect{{#1}}_{{\!#2}}$}}
\newcommand{\vu}[2]{\mbox{$\vect{{#1}}^{{\!#2}}$}}

\begin{document}

\title{Simulations of the dynamic switching of vortex chirality in magnetic nanodisks by a uniform field pulse}

\author{Roman Antos$^{1}$}
\email{antos@karlov.mff.cuni.cz}
\author{Yoshichika Otani$^{2,3}$}

\affiliation{$^{1}$Institute of Physics, Faculty of Mathematics and Physics, Charles University, 12116 Prague, Czech Republic}%
\affiliation{$^{2}$Institute for Solid State Physics, University of Tokyo, Kashiwa 277-8581, Japan}%
\affiliation{$^{3}$RIKEN-ASI, 2-1 Hirosawa, Wako, Saitama 351-0198, Japan.}

\date{\today}

\begin{abstract}
We present a possibility to switch the chirality of a spin vortex occurring in a magnetic nanodisk by applying a uniform in-plane field pulse, based on optimizing its strength and duration. The related spin-dynamical process, investigated by micromagnetic simulations, consists of several stages. After applying the field, the original vortex is expelled from the disk, after which two C-shaped states oscillate between each other. The essence of the method is based on turning the field off at a suitably chosen moment for which the orientation of the C-state will evolve into the nucleation of a vortex with the desirable chirality. This idea simply uses the information about the original chirality present inside the nanodisk during the dynamic process before losing it in saturation, and can thus be regarded as analogous to the recent studies on the polarity switching.
\end{abstract}

\pacs{75.40.Gb, 75.40.Mg, 75.60.Jk, 75.75.+a}
\maketitle

Recent progress in technologies of preparing confined nanomagnetic geometries~\cite{Skomski_JP-CM_2003,Martin_JMMM_2003} has considerably improved the parameters of spin-related devices and shifts the challenges in nanomagnetism towards new approaches.~\cite{Bader_RMP_2006,Srajer_JMMM_2006} Several interesting possibilities emerged from extensive research on spin vortices occurring at equilibrium of flat ferromagnetic cylinders and analogous elements,~\cite{Antos_JPSJ_2008,Hubert_book} whose two binary characteristics, the chirality $c=\pm1$ (counterclockwise (CCW) or clockwise (CW) vortex's flow) and the polarity $p=\pm1$ (up or down orientation of the vortex's out-of-plane polarized core,~\cite{Shinjo_Science_2000,Wachowiak_Science_2002}) determine the dynamic response to ultrafast magnetic field pulses~\cite{Choe_Sci_2004} and are important candidates for nonvolatile magnetic memory~\cite{Nishimura_JAP_2002} and other applications.~\cite{Ross_ARMR_2001, Cowburn_Science_2000}

Switching these states traditionally involved a strong quasistatic magnetic field, out-of-plane for the polarity~\cite{Kikuchi_JAP_2001} and in-plane for the chirality. To switch the latter, however, required a geometric asymmetry, because a saturated symmetric cylinder is---after turning the field off---ordered randomly.~\cite{Taniuchi_JAP_2005, Schneider_APL_2001, Kimura_APL_2007} On the other hand, ultrafast dynamic processes, being forced by field pulses~\cite{Xiao_APL_2006,Hertel_PRL_2007,Waeyenberge_Nature_2006} or electrical currents,~\cite{Yamada_NMAT_2007, Kim_APL_2007, Liu_APL_2007, Sheka_APL_2007, Caputo_PRL_2007} enable to overcome those problems by utilizing consecutively occurring stages. Thus the polarity can be switched via the generation of a vortex--antivortex pair by a short in-plane field pulse of a relatively low amplitude, because the original vortex then annihilates with the antivortex, leaving there the new vortex with the opposite polarity.

We here show, using micromagnetic simulations, that the chirality of a vortex in a symmetric nanodisk can also be switched by an in-plane field pulse, because the information about the original state is still present inside the disk for a certain time (before arriving at the saturated equilibrium), so that a vortex with the opposite chirality can be nucleated after turning the field off at a suitably chosen moment. This method eliminates the need for inhomogeneity of the excitation, which has been introduced by a mask,~\cite{Gaididei_APL_2008} spin transfer torque,~\cite{Choi_APL_2007} or exchange bias.~\cite{Tanase_PRB_2009}

\begin{figure}[b]
\includegraphics[width=8cm]{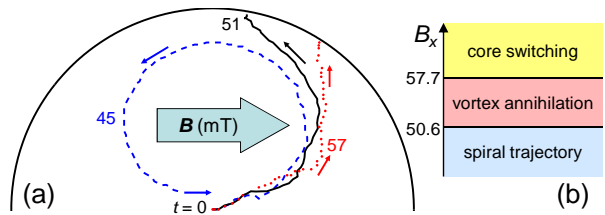}
\caption{\label{fig:trajectory}(color online) Selected trajectories of the vortex core plotted inside the top-viewed contour of the Py nanodisk (a) for applied fields $B_x$ = 45, 51, and 57~mT. For the case of 45~mT (dashed curve) the trajectory corresponds to a spiral motion, whereas for 51 and 57~mT (solid and dotted curves) the trajectory leads the vortex core out of the disk. The chart (b) classifies all processes according to the size of the field.}
\end{figure}

\begin{figure*}[t]
\includegraphics[width=16cm]{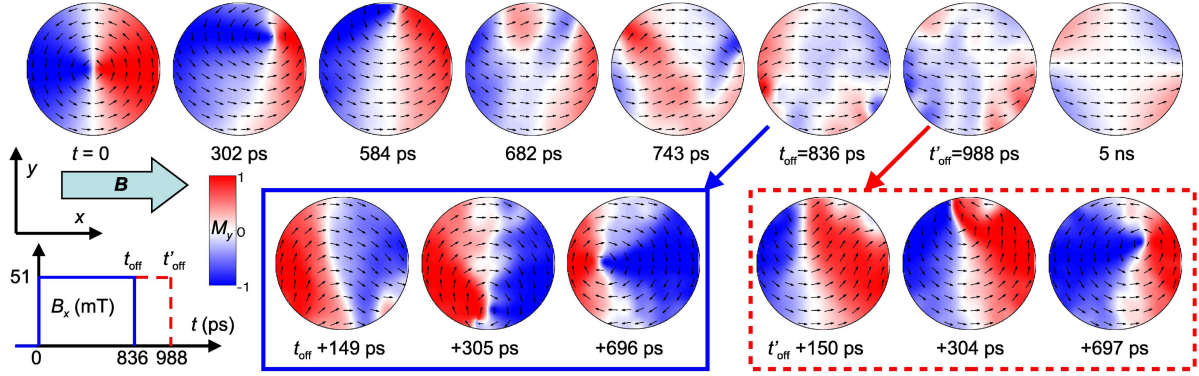}
\caption{\label{fig:vortices}(color online) Time evolution of magnetization distribution after applying a pulse of magnetic field $B_x$ = 51~mT with two examples of temporal duration of 836 and 988~ps. The color scale of~$M_y$, the geometric orientation, and the field pulse are displayed in the bottom-left corner. The top part of the figure demonstrates the evolution during the nonzero field, applied from $t = 0$ until a stable, saturated state at 5~ns. The blue (gray solid) frame displays the evolution after turning the field off at $t_{\rm off}$ = 836~ps (shorter pulse). The red (gray dashed) frame displays the evolution after turning the field off at $t'_{\rm off}$ = 988~ps (longer pulse).}
\end{figure*}

We carry out the time integration of the Landau-Lifshitz-Gilbert (LLG) equation
\begin{equation}\label{eq:LLG}
\frac{\partial\vect{m}}{\partial t} = -\gamma\vect{m}\times\vd{H}{\rm eff} + \alpha\vect{m}\times\frac{\partial\vect{m}}{\partial t},
\end{equation}
where $\vect{m}$ denotes the unit magnetization vector, $\gamma$~the gyromagnetic ratio, $\alpha$~the Gilbert damping parameter, $t$~the time, and $\vd{H}{\rm eff} = -(\mu_0 M_{\rm s})^{-1}\partial E/\partial \vect{m}$ the effective magnetic field, determined from the energy density~$E$ (with $\mu_0$ being the magnetic permeability of vacuum and $M_{\rm s}$ the saturation constant) and being the sum of the externally applied field term~$\vd{H}{\rm ext}$, exchange term $\vd{H}{\rm exch}=(2A/\mu_0M_{\rm s})\nabla^2\vect{m}$ (with $A$ denoting the exchange stiffness constant), and demagnetization term $\vd{H}{\rm d}$. The LLG equation and all the field terms are numerically treated via regular discretization.~\cite{splitting} Thus the exchange field is calculated as $\vd{H}{\rm exch}=(2A/3\mu_0 M_{\rm s}d^2)\sum_j\vu{m}{(j)}$, with the summation over all eight neighbours in the grid.~\cite{Donahue_PhB_1997} The demagnetizing field is generated by surface magnetic charges present on each cell, with help of analytical formulae used for surface integrations as described by Hubert and Sch\"{a}fer.~\cite{Hubert_book} The method was tested to be very close to the OOMMF public code.~\cite{OOMMF_guide} All numerical experiments were carried out on a Py nanodisk of the diameter of 200~nm, thickness of 20~nm, with $c=p=1$, using material parameters $M_{\rm s}=860\:{\rm kAm}^{-1}$, $A=1.3\times10^{-11}\:{\rm Jm}^{-1}$, and $\alpha=0.01$.

Trajectories plotted in Fig.~\ref{fig:trajectory}(a) demonstrate the initiation of motion of the vortex core in the Py nanodisk according to the amplitude of an applied in-plane field pulse. For fields below 50.6~mT, the vortex core starts to move along the field and then traces a CCW spiral around a new equilibrium point, shifted from the disk's center (45~mT chosen as an example). For fields between 50.6 and 57.7~mT, the spiral motion leads the vortex core out of the disk, so that the vortex annihilates. For higher fields a new vortex of the opposite polarity and an antivortex are created, the latter of which annihilates with the old vortex, facilitating thus the switching of the polarity. According to the classification in Fig.~\ref{fig:trajectory}(b), to switch the chirality obviously requires intermediate fields (examples in Fig.~\ref{fig:trajectory}(a) chosen 51 and 57~mT).

The whole process after applying a finite field pulse with the amplitude of 51~mT is demonstrated in Fig.~\ref{fig:vortices}, with two examples of pulse durations of 836 and 988~ps. As already shown in Fig.~\ref{fig:trajectory}, the CCW vortex follows a curve towards the annihilation point at 584~ps, after which a CCW C-like curved state (C-state) appears, which is later transformed into an opposite (CW) C-state, both of which oscillate between each other. According to the snapshots at 682, 743, and 836~ps (and movie1.mov~\cite{video}), the oscillation can be described as alternation between the CCW and CW C-states, the first of which turns into the second via domain wall motion from the north pole to the south pole of the nanodisk. The curvatures of the two C-states continuously decrease, both finally converging into a stable, saturated state (displayed for 5~ns). If we turn the field off at $t_{\rm off}=836$~ps, where the most distinct CW orientation of the C-state is present, then a new vortex nucleates at around $t_{\rm off}+305$~ps with the CW chirality, so that the chirality of the original vortex is switched (the solid frame of Fig.~\ref{fig:vortices} and movie2.mov~\cite{video}). On the other hand, if we turn the field off at $t_{\rm off}=988$~ps, at which the C-state possesses the CCW orientation, the original chirality is preserved (the dashed frame and movie3.mov~\cite{video}). In other words, the orientation of the C-state at which the field is turned off determines the chirality of the nucleated vortex.

\begin{figure}[t]
\includegraphics[width=8cm]{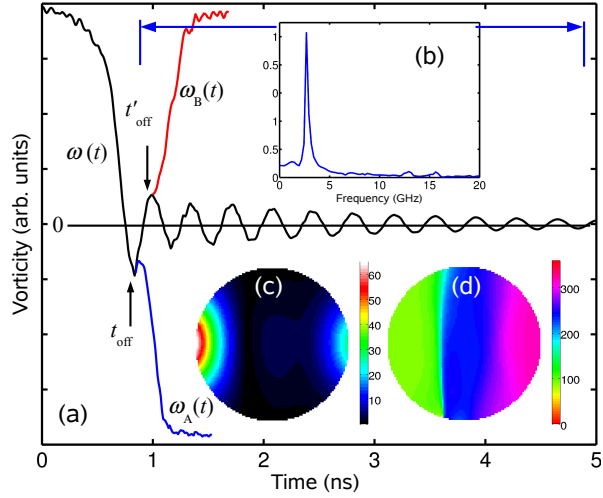}
\caption{\label{fig:vorticity}(color online) Analyzing the dynamic processes. (a) Time evolution of fluid-like vorticity $\omega=[\nabla\times\vect{m}]_z$ (averaged over the nanodisk) after applying the pulse $B_x$ = 51~mT (the $\omega$ curve) and after two examples of turning the field off at $t_{\rm off}$ = 836~ps (the $\omega_{\rm A}$ curve) and $t'_{\rm off}$ = 988~ps (the $\omega_{\rm B}$ curve). The top inset (b) shows the Fourier transform of a selected range of the vorticity function~$\omega(t)$; the selection is depicted by the quotation arrows. The space distribution of magnetization oscillations during the nonzero field for the peak frequency~$\Omega_0$ are displayed in the bottom insets corresponding to the amplitudes (c) and the phases (d) of the oscillations.}
\end{figure}

The dynamic evolution can be also described by a quantity called fluid-like vorticity, $\omega=[\nabla\times\vect{m}]_z$ (with the $z$~axis oriented out-of-plane), whose curves (averaged over the nanodisk) are plotted in Fig.~\ref{fig:vorticity}~(a). Here the positive values represent spin distributions for which the CCW orientation predominates, whereas the negative values represent the CW orientation. Thus the curve $\omega(t)$ exhibits a steep decrease of the vorticity for $t<836$~ps, signifying the disappearance of the CCW vortex and evolution into the CW C-state, and then damped harmonic oscillation signifying the alternation between the CW and CCW C-states. To describe this oscillation in detail, we performed the Fourier transform of a selected part of the vorticity function $\omega(t)$ [Fig.~\ref{fig:vorticity}~(b)], showing a strong frequency peak at $\Omega_0=2.8$~GHz, broadened due to the damping and the space inhomogeneities of the oscillation. For the chosen frequency $\Omega_0$, the spin-dynamical motion in the $x$--$y$ plane can be mathematically described as
\begin{equation}
 \vect{m}(x,y;t)=\vd{m}{0}(x,y)+\vect{a}(x,y)\cos[\Omega_0t + \delta(x,y)],
\end{equation}
where $\vd{m}{0}$ is the immobile component of magnetization, $\vect{a}$ denotes the distribution of amplitudes and $\delta$ the distribution of phases of the oscillations within the nanodisk. These $\vect{a}$ and $\delta$ distributions are displayed in Fig.~\ref{fig:vorticity}~(c,~d), respectively. The highest amplitudes of the oscillations are obviously on the left and right edges of the nanodisk (on the left more significant), while the phases are constant on vertical lines. In other words, the motion is comparable to the flapping of two wings, with a certain asymmetry of the amplitudes and phases due to the asymmetric origin of the vortex annihilation. These modes have been studied in detail previously.~\cite{McMichael_JAP_2005,Gubbiotti_PRB_2005}

For further illustration of the spin evolution after the two pulse durations of 836 and 988~ps (the two bottom frames of Fig.~\ref{fig:vortices}), we also show the corresponding evolutions of the vorticities, $\omega_{\rm A}(t)$ and $\omega_{\rm B}(t)$, plotted in Fig.~\ref{fig:vorticity}~(a), exhibiting the nucleation of the CW or CCW vortex, respectively. Both curves clearly confirm that the negative or positive sign of the vorticity of the alternating C-states uniquely determines the CW or CCW chirality of the final vortex, so that the entire process is controllable by choosing the appropriate pulse length, analogously to the current-driven domain wall motion studied by Thomas et al.~\cite{Thomas_Nat_2006}

\begin{figure}[t]
\includegraphics[width=8cm]{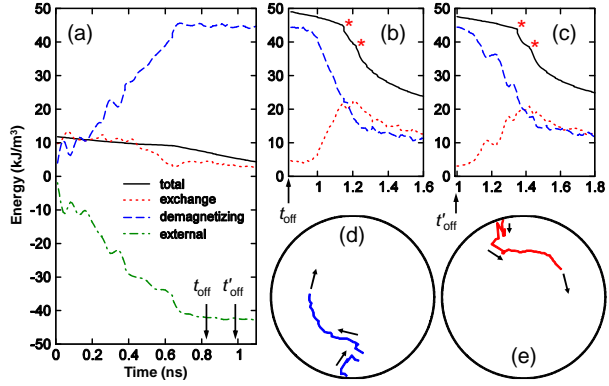}
\caption{\label{fig:energy}(color online) Time evolution of energies. (a) The evolution of the total, exchange, demagnetizing, and external-field energies during the externally applied field $B_x$ = 51~mT. The two quotation arrows show two examples of turning the field off. (b) The evolution of the energies after turning the field off at $t_{\rm off}$ = 836~ps. (c) The analogous evolution for $t'_{\rm off}$ = 988~ps. The trajectories of the cores of nucleated vortices, corresponding to the two cases (b) and (c), are displayed in (d) and (e), respectively.}
\end{figure}

We have also investigated the reliability of the method with respect to the edge roughness~\cite{Nakatani_NMat_2003} by utilizing the deviation from the cylindrical symmetry due to the staircase approximation.~\cite{Garcia-Cervera_JCP_2003, Giovannini_PRB_2004} Thus, besides the excitation field oriented along the $x$-axis, we also performed analogous simulations with fields rotated by 30$^\circ$ and 45$^\circ$ to reveal that the initial process of the vortex expulsion remains nearly same, followed by slightly faster C-state oscillations. However, the time $t_{\rm off}$ discussed above remains quite adequate for the chirality switching, regardless of the field orientation.

To complete our demonstration, Fig.~\ref{fig:energy} displays the micromagnetic energies of the nanodisk, corresponding to all the processes discussed above. Fig.~\ref{fig:energy}~(a) shows the energies during the externally applied field pulse of 51~mT. After applying the field at $t=0$, the vortex gradually changes its distribution to reduce the external field energy, while the demagnetizing energy increases. The exchange energy, which is mostly coupled with the vortex core, significantly vanishes during the vortex annihilation. The total energy exhibits gradual decrease due to damping. Fig.~\ref{fig:energy}~(b) shows the energies after turning the field off at $t_{\rm off}=836$~ps, exhibiting a rapid increase of the exchange energy and a decrease of the demagnetizing energy because a new vortex nucleates. The two red asterisks denote two points of strong release of energy due to the switching of the polarity of the new vortex, accompanied with the creation and annihilation of vortex-antivortex pairs. Fig.~\ref{fig:energy}~(c) shows analogous processes for the case $t'_{\rm off}=988$~ps. Fig.~\ref{fig:energy}~(d,~e) displays the trajectories of the created vortex cores for the two cases corresponding to Fig.~\ref{fig:energy}~(b,~c), respectively.

In summary, we have proposed a switching method for the vortex chirality based on optimizing the strength and duration of the excitation pulse. The method does not require any artificial asymmetry of the nanodisk or any particular field distribution, because the original chirality determines the entire process. The idea is particularly appropriate for nanodisks with relatively small diameters where easy vortex expulsion is possible without core switching. For large diameters, however, the method will also require optimizing the pulse shape or combining the dynamic pulse with quasistatic field. A weak point of the method is the fact that the information about the original polarity is lost after the annihilation of the original vortex, which can be prevented by memorizing the polarity before the vortex is excited and (eventually) by switching it after the new vortex relaxes. Another problem might arise from the fact that the excitation pulse must be applied with a relatively precise duration, which can slightly differ for different shapes of the nanodisk occurring due to fabrication imperfections; however, the minor influence of edge roughness does not cause much trouble.

This work is part of the research plan MSM 0021620834 financed by the Ministry of Education of the Czech Republic and was partially supported by a Marie Curie International Reintegration Grant (no.~224944) within the 7th European Community Framework Programme and by the Grant Agency of the Czech Republic (no.~202/09/P355).


\begin{thebibliography}{37}
\bibitem[{spl()}]{splitting}
\bibinfo{note}{The cylindrical nanodisk of the diameter~$2R$ and thickness~$L$
  uses the splitting of the rectangular area $2R\times 2R\times L$, containing
  the nanodisk, into a regular grid of $N\times N$ rectangular cells of the
  sizes $d_x\times d_y\times L$, where $d_x=d_y=d=2R/N$. We assume $R=100$~nm,
  $L=20$~nm, $N=100$, and $d=2$~nm. Magnetic material fills each $j$th cell
  whose central lateral coordinates~$x_j$,~$y_j$ satisfy the condition
  $x_j^2+y_j^2\leq R$; all the other cells contain air. In the discretized LLG
  equation the unit magnetization $\vect{m}^{(j)}$ is assumed uniform within
  each cell.}
\end{thebibliography}


\end{document}